\documentstyle[11pt,epsfig]{article}

\def\laq{~\raise 0.4ex\hbox{$<$}\kern -0.8em\lower 0.62
ex\hbox{$\sim$}~}
\def\gaq{~\raise 0.4ex\hbox{$>$}\kern -0.7em\lower 0.62
ex\hbox{$\sim$}~}

\begin{document}

\begin{titlepage}
\begin{flushright}
CERN-PH-TH/2008-151
\end{flushright}
\vspace*{1.8 cm}

\begin{center}
\huge
{Stochastic backgrounds of relic gravitons, \\
T$\Lambda$CDM paradigm and the stiff ages}
\vskip1.cm
\large{Massimo Giovannini \footnote{e-mail address: massimo.giovannini@cern.ch}}
\vskip 1.cm
{\it   Department of Physics, Theory Division, CERN, 1211 Geneva 23, Switzerland}
\vskip 0.5cm
{\it  INFN, Section of Milan-Bicocca, 20126 Milan, Italy}
\vskip 1cm

\begin{abstract}
Absent any indirect tests on the thermal history of the Universe prior to the formation of light nuclear elements, 
it is legitimate to investigate situations where, before nucleosyntheis, 
the sound speed of the plasma was larger than  $c/\sqrt{3}$, at most equalling the speed of light $c$. 
In this plausible extension of the current cosmological paradigm, 
hereby dubbed Tensor-$\Lambda$CDM (i.e. T$\Lambda$CDM) scenario, high-frequency gravitons are copiously 
produced. Without conflicting with the bounds on the tensor to scalar ratio stemming from the combined analysis of the three standard cosmological 
data sets (i.e. cosmic microwave background anisotropies, large-scale structure and 
supenovae),  the spectral energy density of the relic gravitons in the T$\Lambda$CDM scenario can be potentially observable by 
wide-band interferometers (in their advanced version) operating  in a frequency window 
which ranges between few Hz and few kHz.
\end{abstract}
\end{center}
\end{titlepage}

\newpage
The only direct informations on the early thermal history of the Universe come, at present, from a background of relic photons which last scattered the electrons at an approximate redshift of $z_{\mathrm{dec}} \simeq 1087$ according to the 5-yr WMAP data release \cite{WMAP51,WMAP52}.   The scrutiny of the Cosmic Microwave Background (CMB)
observables is always conducted within a commonly accepted framework,  i.e. the 
 so-called $\Lambda$CDM paradigm where $\Lambda$ qualifies the dark-energy component (parametrized in terms of a cosmological constant) and  CDM qualifies the (cold) dark matter component. 
The $\Lambda$CDM scenario represents a useful compromise  between the available data and the number of 
ascertainable parameters.  A class of plausible completions 
of the $\Lambda$CDM model contemplates the addition of a post-inflationary phase expanding at a rate which 
is slower than radiation. From the point of view of the fluid properties, the sources generating 
such a dynamics are often called stiff. The spectral energy density 
of the gravitons reentering the Hubble radius during the stiff phase increases with the frequency rather than being nearly constant 
as in the conventional $\Lambda$CDM paradigm. Such an extension requires two parameters: a typical frequency scale, be it 
$\nu_{\mathrm{s}}$ (corresponding to the end of the stiff epoch) and the slope of the spectral energy density 
during the stiff phase.  The supplementary parameters characterizing this scenario (which will be dubbed, in what follows, as 
 T$\Lambda$CDM  for tensor-$\Lambda$CDM) can be determined 
by analyzing the three conventional cosmological data sets 
(i.e. CMB \cite{WMAP51,WMAP52}, large-scale structure \cite{LSS1,LSS2} and supernovae \cite{SN1,SN2}) 
 in conjunction with the forthcoming data of wide-band intereferometers \cite{ligo,virgo,tama,geo}.  At the moment interferometers are only able to provide interesting upper limits on the spectral energy density 
of the relic gravitons \cite{stoch1}. The foreseen sensitivities of the so-called advaced Ligo \cite{ligo} 
 will still be inadequate to probe the relic gravitons produced within the conventional $\Lambda$CDM scenario. 
 Nonetheless the very same sensitivities of the interferometers in their advanced version  will be definitely sufficient to probe directly the parameter space of the T$\Lambda$CDM scenario. 
 
Consider therefore the evolution of the tensor modes in conformally flat 
background geometries which are, incidentally, the ones currently preferred in the context of the 
 $\Lambda$CDM paradigm \cite{WMAP51,WMAP52}.
A conformally flat  background geometry in four space-time dimensions, 
by definition, is characterized by a metric 
 $\overline{g}_{\mu\nu} = a^2(\tau) \eta_{\mu\nu}$ where $\eta_{\mu\nu}$ is the Minkowski metric with signature 
 mostly minus and $\tau$ is the so-called conformal time coordinate. 
 The tensor fluctuations of the geometry are defined with respect to the three-dimensional Eucledian sub-manifold as 
\begin{equation}
\delta_{\mathrm{t}}^{(1)}g_{ij} = - a^2 h_{ij}, \qquad \delta_{\mathrm{t}}^{(1)} g^{ij} = - \frac{h^{ij}}{a^2}, 
\qquad \delta_{\mathrm{t}}^{(2)} g^{i j} = -  \frac{h^{i}_{k} h^{kj}}{a^2},\qquad \partial_{i} h^{i}_{j} = h_{i}^{i}=0,
\label{EQ1}
\end{equation}
 where Latin indices run over the spatial dimensions; 
 $\delta_{\mathrm{t}}^{(1)}$ and $\delta_{\mathrm{t}}^{(2)}$  denote, respectively, the first 
 and second order tensor fluctuations of the corresponding quantity.  Since $h_{ij}$ is  
 a (divergenceless and traceless) rank-two tensor in three spatial dimensions,
  it carries two physical polarizations. Defining 
 three mutually orthogonal directions as $\hat{k}_{i} = k_{i}/|\vec{k}|$,  $\hat{m}_{i} = m_{i}/|\vec{m}|$ and
 $\hat{n}_{i} =n_{i}/|\vec{n}|$, the two polarizations of the gravitons in a conformally flat background are nothing but  
 \begin{equation}
 \epsilon_{ij}^{(\oplus)}(\hat{k}) = (\hat{m}_{i} \hat{m}_{j} - \hat{n}_{i} \hat{n}_{j}), \qquad 
  \epsilon_{ij}^{(\otimes)}(\hat{k}) = (\hat{m}_{i} \hat{n}_{j} + \hat{n}_{i} \hat{m}_{j}), \qquad  \epsilon_{ij}^{(\lambda)} \epsilon_{ij}^{(\lambda')} = 2 \delta_{\lambda\lambda'}.
 \label{EQ2}
 \end{equation}
By perturbing the Einstein-Hilbert action to second order in the tensor amplitude $h_{ij}$, the action for the gravitons can be written, up to total derivatives, as 
\begin{equation}
S_{\mathrm{gw}} = \delta^{(2)}_{\mathrm{t}} S = \frac{1}{8\ell_{\mathrm{P}}^2} \int d^{4} x \sqrt{- \overline{g}} \,\,\overline{g}^{\mu\nu} \,\,\partial_{\mu} h_{ij} \partial_{\nu} h^{ij},\qquad \ell_{\mathrm{P}} = \sqrt{8\pi G} =\frac{8\pi}{M_{\mathrm{P}}}= 
\frac{1}{\overline{M}_{\mathrm{P}}}.
\label{EQ3}
\end{equation}
Up to a rescaling of the amplitude in terms of the Planck length, the canonical normal modes of the action (\ref{EQ3}) 
 are given   by $\mu_{ij} = a h_{ij}$  since, in a  conformally flat background, $\sqrt{-\overline{g}}\,\, \overline{g}^{\mu\nu} \,\,\to 
 a^2(\tau) \eta^{\mu\nu}$. The mode expansion of the canonical field operator is thus given by:
 \begin{equation}
 \hat{\mu}_{ij}(\vec{x},\tau)= \frac{\sqrt{2} \ell_{\mathrm{P}}}{(2\pi)^{3/2}} \sum_{\lambda}
  \int d^{3}k\,\, \epsilon_{ij}^{(\lambda)}(\hat{k})\, 
 \biggl[ \hat{a}_{\vec{k},\lambda} \,f_{k,\lambda}(\tau) e^{- i \vec{k} \cdot \vec{x}} +  \hat{a}_{\vec{k},\lambda}^{\dagger}\, f_{k,\lambda}^{*}(\tau) 
 e^{ i \vec{k} \cdot \vec{x}}\biggr],
 \label{EQ4}
 \end{equation}
 where $[\hat{a}_{\vec{k},\lambda}, \hat{a}^{\dagger}_{\vec{p},\lambda'}] = \delta^{(3)}(\vec{k} - \vec{p})
\delta_{\lambda\lambda'}$. It will be hereby assumed that the field operators are in the 
vacuum at the onset of the inflationary evolution. Thus the initial state $|0\rangle$ 
(annihilated by $\hat{a}_{\vec{k},\lambda}$) minimizes 
the tensor Hamiltonian when all the wavelengths of the field are shorter than the event horizon at the onset of the inflationary evolution (see, for instance, \cite{THTool}). In Eq. (\ref{EQ4}) $f_{k,\lambda}$ are the (complex) tensor mode functions obeying
\begin{equation}
f'_{k,\lambda} = g_{k,\lambda}, \qquad g_{k,\lambda}' = - [ k^2 - ({\mathcal H}' + {\mathcal H}^2)] f_{k,\lambda},\qquad 
{\mathcal H} = \frac{a'}{a},
\label{EQ5}
\end{equation}
where the prime denotes a derivation with respect to the conformal time coordinate $\tau$.
Defining $p_{\mathrm{t}}$ and $\rho_{\mathrm{t}}$ as the total pressure and as the total energy density 
of the plasma, the Friedmann-Lema\^itre equations read:
\begin{equation}
{\mathcal H}^2 = \frac{8\pi G}{3} a^2 \rho_{\mathrm{t}}, \qquad {\mathcal H}^2 - {\mathcal H}' = 4\pi G a^2 (\rho_{\mathrm{t}} + p_{\mathrm{t}}), \qquad \rho_{\mathrm{t}}' + 3 {\mathcal H}(\rho_{\mathrm{t}} + p_{\mathrm{t}}) =0.
\label{EQ6}
\end{equation}
If the inflationary phase is  
suddenly followed by the radiation-dominated phase,  the energy-density of the inflaton is instantaneously converted into a radiation. This approximation is customarily employed to assess the number of inflationary e-folds \cite{THTool,efolds}.
Given our ignorance on the thermal history of the plasma prior to nucleosynthesis, 
the inflationary phase might  not be suddenly followed by the radiation dominated phase \cite{MG2,MG3}.
Provided the transition between inflation and radiation is sufficiently stiff 
(and long)  high-frequency gravitons can be 
copiously produced \cite{MG2,MG3}. A relativistic plasma is said to be stiff if its sound speed 
is larger than the sound speed of a gas of ultra-relativistic particles \footnote{Natural units $\hbar= c= k_{\mathrm{B}} =1$ are used throughout the script.} i.e.  $1/\sqrt{3}$. The total sound speed and the barotropic index are defined, respectively, as:
 \begin{equation}
c_{\mathrm{st}}^2 = \frac{\partial p_{\mathrm{t}}}{\partial \rho_{\mathrm{t}}}
=w_{\mathrm{t}} - \frac{1}{3} \frac{\partial (w_{\mathrm{t}} + 1)}{\partial\ln{a}}, \qquad w_{\mathrm{t}} = \frac{p_{\mathrm{t}}}{\rho_{\mathrm{t}}},
\label{EQ7}
\end{equation}
where, in the second equality defining $c_{\mathrm{st}}^2$, Eq. (\ref{EQ6}) has been used.
 In the primeval plasma, stiff phases can arise: this idea goes back to the pioneering suggestions 
of Zeldovich \cite{ZEL1} in connection with the entropy problem.  If an inflationary phase precedes 
a stiff phase the spectral energy density of the relic gravitons increases with frequency and the 
 typical length of the stiff epoch can be  determined by back-reaction effects  \cite{MG2}.  
 In \cite{MG3} the techniques of \cite{MG2} were applied 
 to assess the spectral energy density in the models of quintessential inflation which were 
 developed in \cite{PV1}. There were various reprises of these ideas (see, for instance, \cite{REP} and references therein). 
 A (causal) upper limit on $w_{\mathrm{t}}$ and $c_{\mathrm{st}}$ is the speed of light, i.e. 
 $w_{\mathrm{t}} \leq c_{\mathrm{st}} \leq 1$ \cite{EMM}. 
 
 Collisionless species couple to the tensor modes of the geometry. Defining as $\Pi_{ij}$ the anisotropic 
stress of the plasma we will actually have that below temperatures ${\mathcal O}(\mathrm{MeV})$, i.e. 
after weak interactions fall out of thermal equilibrium, the evolution equations for the classical 
amplitude corresponding to the quantum operators  of Eq. (\ref{EQ4}) reads 
\begin{equation}
\mu_{ij}'' - \nabla^2 \mu_{ij} - ({\mathcal H}' + {\mathcal H}^2) \mu_{ij} = - 16\pi G a^3 \Pi_{ij}.
\label{EQ8}
\end{equation}
The coupling to the anisotropic stress induces computable differences 
on the spectral energy density of the relic gravitons.  
The effects of neutrino free streaming has been investigated both semi-analytically and numerically in \cite{nu1} (see 
also \cite{TR1,TR2,TR3}). With  this caveat on collisionless species, 
 Eqs. (\ref{EQ5})--(\ref{EQ6}) can be solved numerically;
  the spectral energy density of the relic gravitons (and the related power spectrum) can then be assessed. 

The  definition of the energy-momentum  pseudo-tensor 
of the gravitational field always involves a certain degree of ambiguity. After getting rid of the tensor structure 
by making explicit the two physical polarizations,
the action of Eq. (\ref{EQ3}) is just the action of two minimally coupled scalar fields in a conformally 
flat geometry of Friedmann-Robertson-Walker (FRW) type. The energy-momentum 
pseudo-tensor of relic gravitons in a FRW background just given by \cite{ford}
\begin{equation}
T_{\mu}^{\nu} = \frac{1}{4 \ell_{\mathrm{P}}^2} \biggl[ \partial_{\mu} h_{ij} \partial^{\nu} h^{ij} - \frac{1}{2} \delta_{\mu}^{\nu} \overline{g}^{\alpha\beta} \partial_{\alpha} h_{ij} \partial_{\beta}h^{ij} \biggr] = \frac{1}{2\ell_{\mathrm{P}}^2} \sum_{\lambda} \biggl[ \partial_{\mu} h_{(\lambda)}
 \partial^{\nu} h^{(\lambda)} - \frac{1}{2} \overline{g}^{\alpha\beta} \partial_{\alpha} h_{(\lambda)}
 \partial_{\beta} h_{(\lambda)} \delta_{\mu}^{\nu}\biggr],
\label{EQ9}
\end{equation}
where the second equality follows from the first by using $h_{ij} = \sum_{\lambda} h_{(\lambda)} \epsilon_{ij}^{\lambda}$ 
and by recalling the orthogonality condition appearing in Eq. (\ref{EQ2}). 
 In a complementary perspective \cite{isaacson}, the energy-momentum pseudo-tensor is instead defined from the second-order fluctuations of the Einstein tensor, i.e. 
\begin{equation}
{\mathcal T}_{\mu}^{\nu} = - \frac{1}{\ell_{\mathrm{P}}^2} \delta^{(2)}_{\rm t} {\cal G}_{\mu}^{\nu},\qquad 
{\mathcal G}_{\mu}^{\nu} = R_{\mu}^{\nu} - \frac{1}{2} \delta_{\mu}^{\nu} R,
\label{EQ10}
\end{equation}
where the superscript at the right hand side denotes the second-order 
fluctuation of the corresponding quantity while the subscript refers to the tensor nature of the fluctuations.  The two 
definitions seem very different but the energy densities and pressures derived in the two approaches give coincident 
 results as soon as the corresponding  wavelengths are inside the Hubble radius, i.e. $k > {\mathcal H}$.   In the opposite limit Eqs. (\ref{EQ9})--(\ref{EQ10})   seem superficially different but give consistent quantitative results once they are compared on a particular background geometry \cite{BR}. 
 
By definition, $\rho_{\mathrm{GW}}(\vec{x},\tau) = \langle 0| T_{0}^{0}(\vec{x},\tau)|0 \rangle$ 
where $|0\rangle$ is, again, the state annihilated by $a_{k,\lambda}$.
Recalling that the mode functions of each 
polarization coincide ( i.e., in Eq. (\ref{EQ5}), $f_{k,\oplus}= f_{k,\otimes} = f_{k}$ and analogously for $g_{k}$)
the spectral energy density in critical units can then be expressed as:
\begin{eqnarray}
&& \Omega_{\mathrm{GW}}(k,\tau) = \frac{1}{\rho_{\mathrm{crit}}} \frac{d \rho_{\mathrm{GW}}}{d \ln{k}} = 
\frac{k^3}{2\,\pi^2\,a^4\,\rho_{\mathrm{crit}}} \Delta_{\rho}(k,\tau),\qquad \rho_{\mathrm{crit}}= \frac{3 H^2}{8\pi G} = 3 \overline{M}_{\mathrm{P}}^2 H^2,
\nonumber\\ 
&& \Delta_{\rho}(k,\tau)= \biggl\{ |g_{k}(\tau)|^2 + ( k^2 + {\mathcal H}^2) |f_{k}(\tau)|^2
- {\mathcal H}[ f_{k}^{*}(\tau)g_{k}(\tau) + f_{k}(\tau) g_{k}^{*}(\tau)] \biggr\}.
\label{EQ11} 
\end{eqnarray}
The spectral energy density of the relic gravitons can be related to the power spectrum which is, by definition,  
the Fourier transform of the two-point function evaluated at equal times, i.e. using Eq. (\ref{EQ4})
\begin{equation}
\langle 0| \hat{h}_{ij}(\vec{x},\tau) \hat{h}^{ij}(\vec{y},\tau) |0\rangle = \int d\ln{k} {\mathcal P}_{\mathrm{T}}(k,\tau) 
\frac{\sin{k r}}{kr}, \qquad {\mathcal P}_{\mathrm{T}}(k,\tau) = 4\ell_{\mathrm{P}}^2\frac{k^3}{\pi^2 a^2(\tau)} |f_{k}(\tau)|^2,
\label{EQ12}
\end{equation}
where $r = |\vec{x} - \vec{y}|$. Quantum fluctuations present during the inflationary phase are amplified 
with nearly scale-invariant slope. The inflationary power spectra are then parametrized in terms 
of the tensor and scalar spectral indices, i.e., respectively, $n_{\mathrm{T}}$ and $n_{\mathrm{s}}$:
\begin{equation}
r_{\mathrm{T}} = \frac{{\mathcal A}_{\mathrm{T}}}{{\mathcal A}_{{\mathcal R}}}, \qquad 
\overline{{\mathcal P}}_{\mathrm{T}}(k) = {\mathcal A}_{\mathrm{T}} 
\biggl(\frac{k}{k_{\mathrm{p}}}\biggr)^{n_{\mathrm{T}}}, \qquad 
\overline{{\mathcal P}}_{\mathcal R}(k) = {\mathcal A}_{\mathcal R} \biggl(\frac{k}{k_{\mathrm{p}}}\biggr)^{n_{\mathrm{s}}-1},
\label{EQ13}
\end{equation}
where $k_{\mathrm{p}} = 0.002\,\, \mathrm{Mpc}^{-1}$ is the so-called pivot wave-number which corresponds to an effective 
multipole $\ell_{\mathrm{eff}}\simeq 30$. In the context of the $\Lambda$CDM paradigm, the 5-yr WMAP data 
alone imply ${\mathcal A}_{{\mathcal R}} = 2.41 \times 10^{-9}$ (slightly different values can be 
obtained if different data sets are combined but these differences do not affect the features addressed here). 
The tensor amplitude is therefore estimated 
by setting limits on $r_{\mathrm{T}}$ which is, by definition, the ratio between the tensor and the scalar amplitudes evaluated 
at the pivot scale $k_{\mathrm{p}}$.
The inferred upper bounds on $r_{\mathrm{T}}$, range from $r_{\mathrm{T}} < 0.2$ (in the case of the WMAP 5-yr data 
alone \cite{WMAP51,WMAP52}) up to $r_{\mathrm{T}} < 0.43$ when the WMAP 5-yr data are combined 
with the large-scale \cite{LSS1,LSS2} and supernova data \cite{SN1,SN2} (see also the thorough analyses 
reported in \cite{WMAP51,WMAP52}).   In the minimal version of the inflationary dynamics the tensor spectral slope (i.e. $n_{\mathrm{T}}$),  the slow roll parameter $\epsilon$ as well as $r_{\mathrm{T}}$ are all related:
\begin{equation}
n_{\mathrm{T}} \simeq - \frac{r_{\mathrm{T}}}{8} \simeq - 2 \epsilon,\qquad 
\epsilon = - \frac{\dot{H}}{H^2}>0,
\label{EQ14}
\end{equation}
where $\epsilon$ measures, as indicated, the (slight) decrease of the Hubble $H$ rate during 
the quasi-de Sitter phase of expansion and the overdot in the last equation denotes a derivation 
with respect to the cosmic time coordinate.  
Usually Eq. (\ref{EQ12}) is computed at the 
present epoch and then, in a second step, the spectral energy density of the relic gravitons is derived \cite{FS1} 
(see also, for instance, \cite{TR1,TR2,TR3}).  The spectral energy density can be  also directly assessed by numerical means without 
passing through the transfer function of the amplitude: this will be the approach followed here.  
Within the first strategy the power spectrum is given by
\begin{equation}
{\mathcal P}_{\mathrm{T}}(k,\tau_{0}) = \frac{9\,j^2_{1}(k\tau_{0})}{|k\tau_{0}|^2}
 \biggl[ 1 + c_{1} \biggl(\frac{k}{k_{\mathrm{eq}}}\biggr) + b_{1}\biggl(\frac{k}{k_{\mathrm{eq}}}\biggr)^2\biggr] \overline{{\mathcal P}}_{\mathrm{T}}(k) , 
\label{EQ15}
\end{equation}
where\footnote{By repeating the analysis of \cite{TR1} we obtained $a_{1}= 1.260$ and $b_{1}= 2.683$ 
which is fully compatible with the results of \cite{TR1}. In the approach of \cite{TR1} the calculation of the amplitude transfer function, in fact, involve a delicate matching on the phases of the tensor mode 
functions. Conversely, if the transfer function is computed directly for the spectral energy density, the oscillatory contributions are suppressed as the wavelengths get shorter than the Hubble radius (see below).},
 according to \cite{TR1}, $c_{1} = 1.34$ and $b_{1} = 2.50$.
In Eq. (\ref{EQ15}) $j_{1}(y) = (\sin{y}/y^2 - \cos{y}/y)$ is the spherical Bessel function of first kind which is related to the approximate solution of the evolution equations for the tensor mode functions whenever the solutions  are computed deep in the matter-dominated phase (i.e. $a(\tau) \simeq \tau^2$).  To obtain the spectral energy density, Eq. (\ref{EQ11})  must then be evaluated in the limit 
$k^2 \gg {\mathcal H}^2$ (i.e. wavelengths inside the Hubble radius). In the latter limit the tensor mode functions 
satisfy $|g_{k}(\tau)| \simeq k f_{k}(\tau)$ and Eq. (\ref{EQ11}) then gives:
\begin{equation}
\Omega_{\mathrm{GW}}(k,\tau_{0}) = \frac{k^2}{ 12 {\mathcal H}_{0}^2} {\mathcal P}_{\mathrm{T}}(k,\tau_{0}), \qquad 
\lim_{k \gg k_{\mathrm{eq}}} \Omega_{\mathrm{GW}}(k,\tau_{0})  \simeq \frac{3 b_{1}}{8 a_{0}^2 H_{0}^2 \tau_{0}^4 k_{\mathrm{eq}}^2} \biggl(\frac{k}{k_{\mathrm{p}}}\biggr)^{n_{\mathrm{T}}}.
\label{EQ16}
\end{equation}
Since ${\mathcal P}_{\mathrm{T}}(k,\tau_{0})$ oscillates also $\Omega_{\mathrm{GW}}(k,\tau_{0})$ will oscillate. 
In the limit $k \gg k_{\mathrm{eq}}$ 
the cosine will dominate the expression of $j_{1}(k\tau_{0})$and the second result of Eq. (\ref{EQ16}) arises by replacing$\cos^2({k\tau_{0}}) \to 1/2$. 
If we take $b_{1} = 2.5$ in the second relation of Eq. (\ref{EQ16}), then $3 b_{1}/8 \equiv 15/16 = 0.9375 $. If 
we take instead our results (i.e. $b_{1} = 2.683$) we will get, for the same quantity, $1.006$. What appears in Eq. (\ref{EQ15}) 
 is the transfer function of the tensor amplitude which literally transfers    
the power spectrum $\overline{{\mathcal P}}_{\mathrm{T}}$ inside the Hubble radius. 

\begin{figure}
\begin{center}
\begin{tabular}{|c|c|}
      \hline
      \hbox{\epsfxsize = 7.2 cm  \epsffile{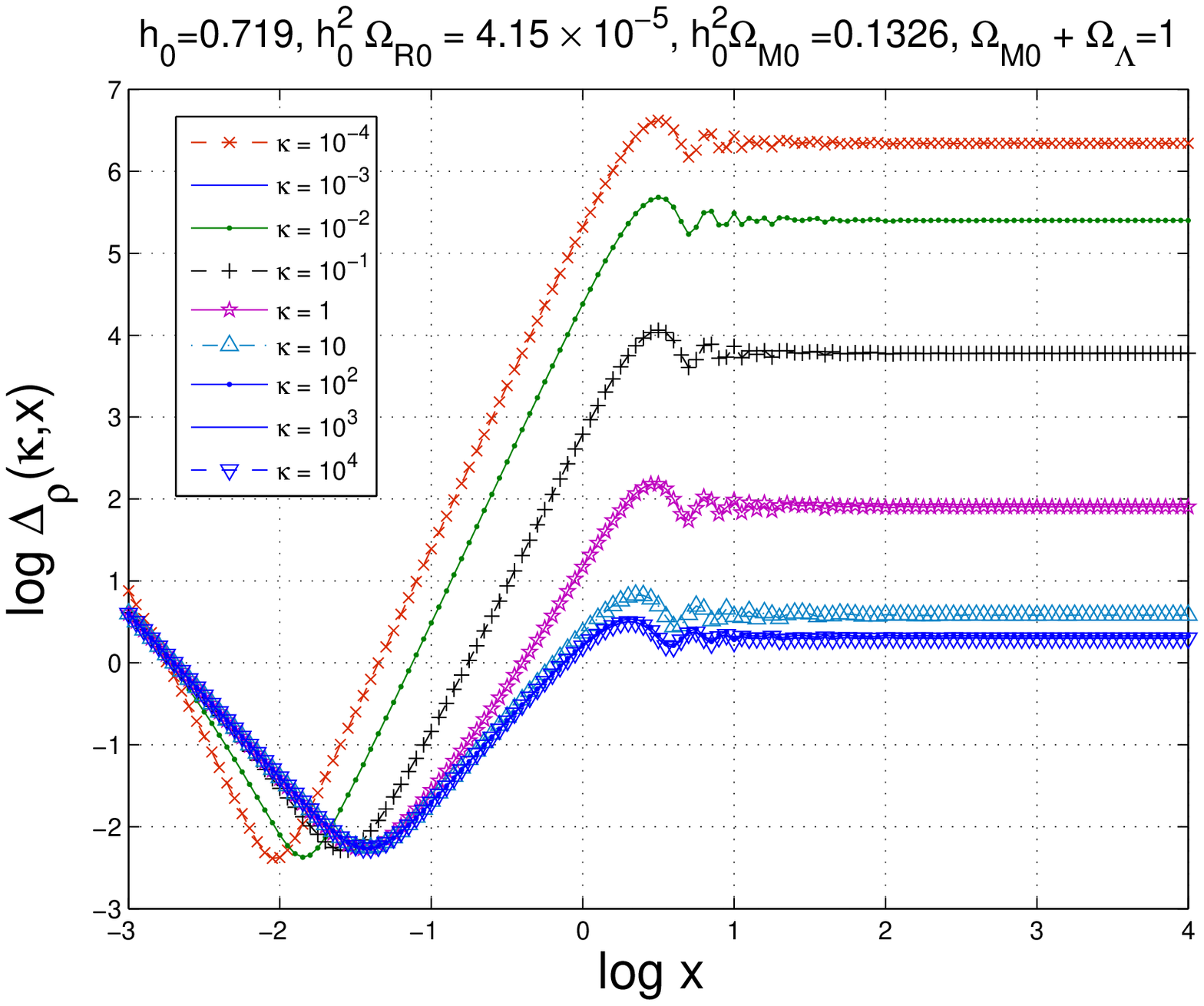}} &
      \hbox{\epsfxsize = 7.2 cm  \epsffile{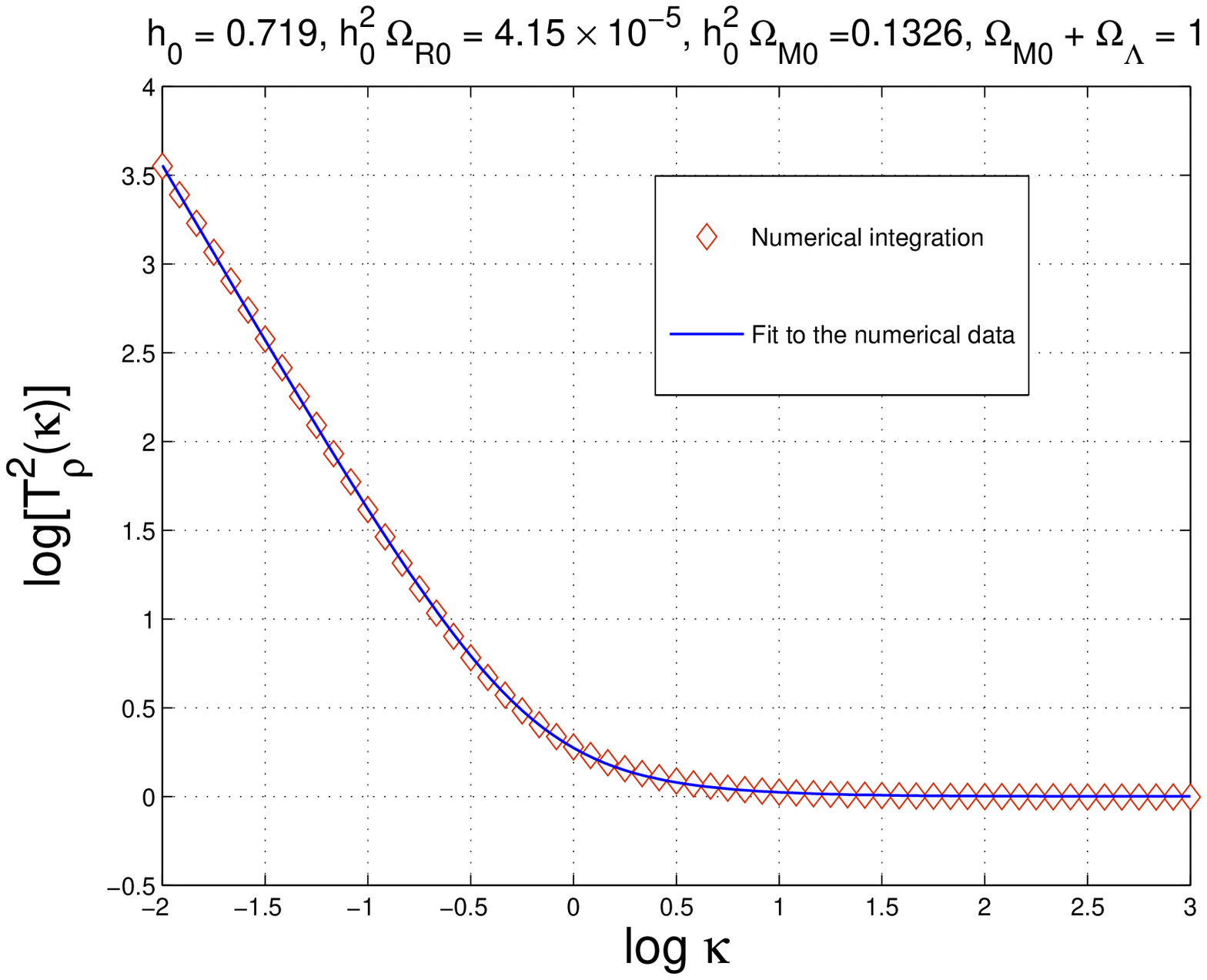}}\\
      \hline
\end{tabular}
\end{center}
\caption{The spectral energy density (see Eq. (\ref{EQ11})) is integrated across the radiation matter transition 
for different values of $\kappa = k/k_{\mathrm{eq}}$ (plot at the left). A finer  grid in $\kappa$ (plot at the right), allows for the computation of the energy transfer function 
whose form can be fitted by an approptiate analytical expression (see Eq. (\ref{EQ19}).} 
\label{FIGURE1}
\end{figure}

In a complementary perspective, the consistent numerical solution of Eqs. (\ref{EQ5})--(\ref{EQ6}) allows for 
a numerical calculation of $\Omega_{\mathrm{GW}}(k,\tau)$ according to Eq. (\ref{EQ11}). 
Instead of fitting the final result in terms of a putative (semi-analytic) amplitude for the mode function, 
the momentum (or frequency) profile of the spectral energy density will be obtained directly by numerical 
methods. As usual, initial conditions for the numerical integration
are given  for $k\tau \ll 1$. The system is then followed through Hubble crossing (i.e. $k\tau \simeq 1$). 
Finally, when $k\tau \gg 1$ the expression of $\Delta_{\rho}(k,\tau)$ can be  read-off 
in the asymptotic regime. In Fig. \ref{FIGURE1}  (plot at the left) the numerical integration across 
the radiation-matter transition is illustrated.  Instead of phrasing the numerical integration in terms 
of $k$ and $\tau$, it is practical to use $x = k\tau$ and $\kappa = k/k_{\mathrm{eq}}$ as preferred variables.
 To be accurate on the initial conditions a fully analytic solution of Eq. (\ref{EQ6}), valid 
 across the radiation-maatter transition,  can be safely employed:
 \begin{eqnarray}
 && a(\tau) = a_{\mathrm{eq}}\biggl[ \biggl(\frac{\tau}{\tau_{1}}\biggr)^2 + 2 \biggl(\frac{\tau}{\tau_{1}}\biggr)\biggr],\qquad 
 \frac{a_{0}}{a_{\mathrm{eq}}} = 1 + z_{\mathrm{eq}} = 3195.17 \,  \biggl(\frac{h_{0}^2 \Omega_{\mathrm{M}0}}{0.1326}\biggr)
 \biggl(\frac{h_{0}^2 \Omega_{\mathrm{R}0}}{4.15 \times 10^{-5}}\biggr)^{-1},
 \nonumber\\
&& \tau_{\mathrm{eq}} = (\sqrt{2} -1)\tau_{1} = 120.658 \biggl(\frac{h_{0}^2 \Omega_{\mathrm{M}0}}{0.1326}\biggr)^{-1} \biggl(\frac{h_{0}^2 \Omega_{\mathrm{R}0}}{4.15 \times 10^{-5}}\biggr)^{1/2}\,\,\mathrm{Mpc}.
 \label{EQ17}
 \end{eqnarray}
In the limit  $x \ll 1$ the initial conditions for the mode functions are determined directly (and up to phase factors) 
from Eq. (\ref{EQ13}). Since the system is linear, the tensor mode functions  can be always rescaled 
through their initial value; the energy transfer function is therefore defined by the following limit 
\begin{equation}
\lim_{x \gg 1} \Delta_{\rho}(\kappa,x) \equiv T^{2}_{\rho}(\kappa) \Delta_{\rho}(\kappa,x_{\mathrm{i}}), \qquad 
x_{\mathrm{i}} \ll 1.
\label{EQ18}
\end{equation}
In Fig. \ref{FIGURE1} the results of the numerical integration are reported in terms of $\Delta_{\rho}(\kappa,x)$ 
for different values of $\kappa$ (see plot at the left). Always in Fig. \ref{FIGURE1} (plot at the right),
  $T^{2}_{\rho}(\kappa)$ can be 
computed numerically: the diamonds correspond to the numerical points and the full line (in the plot at the right)
is the numerical fit obtained by means of standard methods in the analysis of the regressions: 
\begin{equation}
T_{\rho}(k/k_{\mathrm{eq}}) = \sqrt{1 + c_{2}\biggl(\frac{k_{\mathrm{eq}}}{k}\biggr) + b_{2}\biggl(\frac{k_{\mathrm{eq}}}{k}\biggr)^2},\qquad c_{2}= 0.5238,\qquad
b_{2}=0.3537.
\label{EQ19}
\end{equation}
For a successful numerical determination of $T_{\rho}(\kappa)$ the initial integration 
variable should be sufficiently small (i.e. $x_{\mathrm{i}} = k\tau_{\mathrm{i}} \ll1$) in such a way that, at the 
initial time, the mode $k_{\mathrm{eq}} = \tau_{\mathrm{eq}}^{-1}$  had a corresponding 
wavelength much smaller than the Hubble radius at $\tau_{\mathrm{i}}$. Second, $x_{\mathrm{f}}$ should be 
sufficiently large so that, effectively, $\Delta_{\rho}(x_{\mathrm{f}}, \kappa)$ is constant up to terms 
${\mathcal O}(1/x_{\mathrm{f}})$ (see also below Eq. (\ref{EQ26})). Finally, the grid in $\kappa$ should be sufficiently fine to allow for a reasonable fit. 
Using Eq. (\ref{EQ19}), the spectral energy density can be written, in the absence of free streaming, as
\begin{equation}
h_{0}^2 \Omega_{\mathrm{GW}}(\nu,\tau_{0}) = {\mathcal N}_{\rho}  T^2_{\rho}(\nu/\nu_{\mathrm{eq}}) r_{\mathrm{T}} \biggl(\frac{\nu}{\nu_{\mathrm{p}}}\biggr)^{n_{\mathrm{T}}} e^{- 2 \beta \frac{\nu}{\nu_{\mathrm{max}}}}, \qquad 
{\mathcal N}_{\rho} = 4.165 \times 10^{-15} \biggl(\frac{h_{0}^2 \Omega_{\mathrm{R}0}}{4.15\times 10^{-5}}\biggr),
\label{EQ20}
\end{equation}
where $\beta=6.33$ has been determined numerically assuming a smooth transition 
between inflation and radiation \cite{MG5}. 
Equation (\ref{EQ20}), unlike  Eqs. (\ref{EQ15})--(\ref{EQ16}), is not strongly oscillating.  The rationale for this difference 
 is that, when computing $\Delta_{\rho}(\kappa,x)$, the  oscillating contributions get 
 dynamically suppressed as the wavelengths get shorter than the Hubble radius. A way 
 of understanding this effect is to notice that the crudest approximation for the mode functions in the limit 
 $k\tau \gg 1$ are simple plane waves, i.e. 
 \begin{equation}
\overline{f}_{k}(\tau) = \frac{1}{\sqrt{2 k}} \biggl[ c_{+}(k) e^{-i k \tau} + c_{-}(k) e^{i k\tau}\biggr],\qquad 
\overline{g}_{k}(\tau) = - i \sqrt{\frac{k}{2}} \biggl[ c_{+}(k) e^{-i k\tau} - c_{-}(k) e^{i k\tau}\biggr].
\label{EQ24}
\end{equation}
Using Eq. (\ref{EQ24}) into Eq. (\ref{EQ11}) and enforcing the limit $x \to x_{\mathrm{f}} \gg1$, 
\begin{equation}
 \Delta_{\rho}(\kappa,x_{\mathrm{f}})  =
 \kappa ( |c_{+}(\kappa)|^2 + |c_{-}(\kappa)|^2)  + {\mathcal O}\biggl(\frac{1}{x_{\mathrm{f}}}\biggr),
 \label{EQ26}
 \end{equation}
which proofs that the oscillating contributions are suppressed and that $\Delta_{\rho}(\kappa,x_{\mathrm{f}})$ is 
proportional to what  are called, in the jargon, mixing coefficients. 
\begin{figure}
\begin{center}
\begin{tabular}{|c|c|}
      \hline
      \hbox{\epsfxsize = 7.2 cm  \epsffile{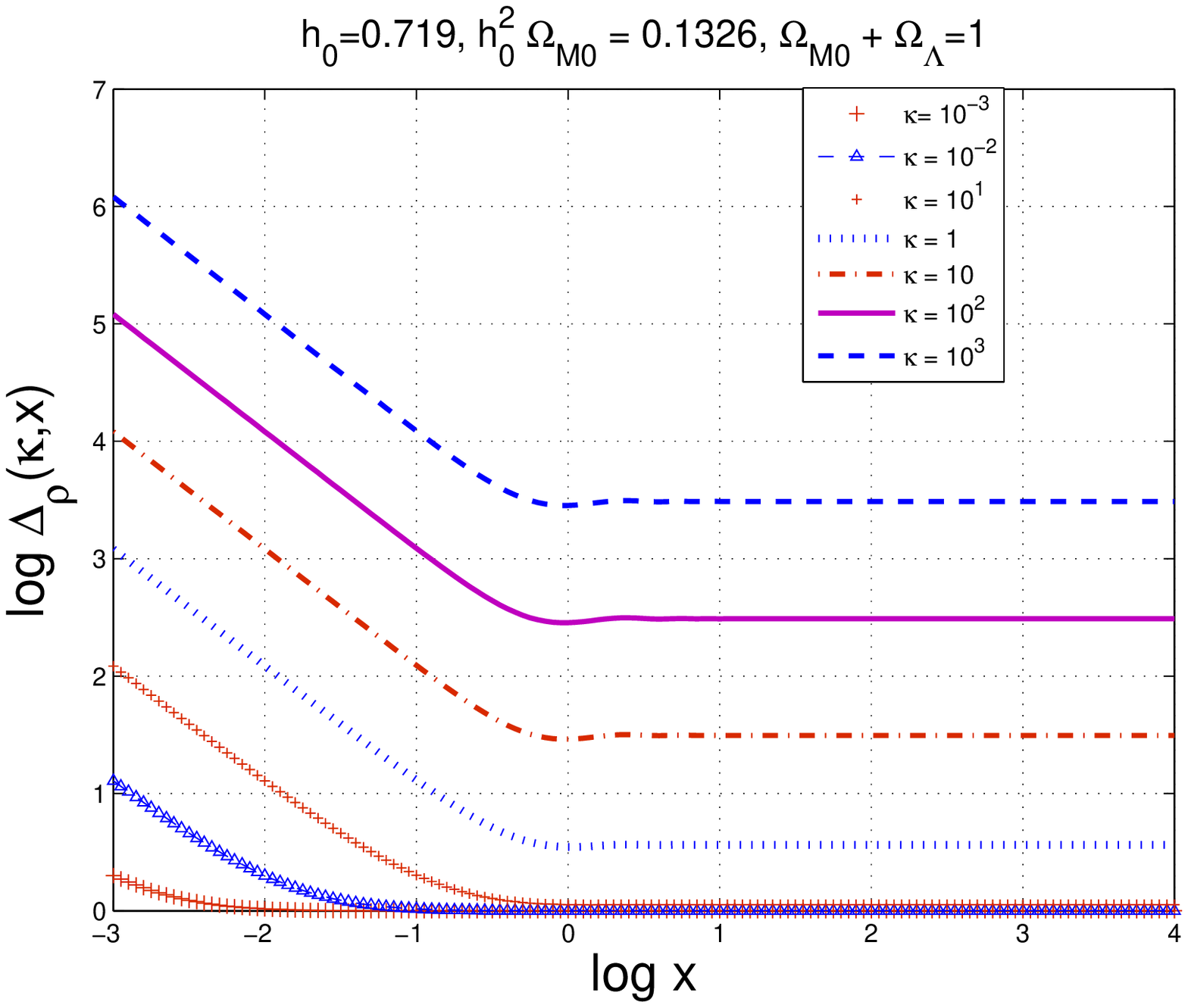}} &
      \hbox{\epsfxsize = 7.2 cm  \epsffile{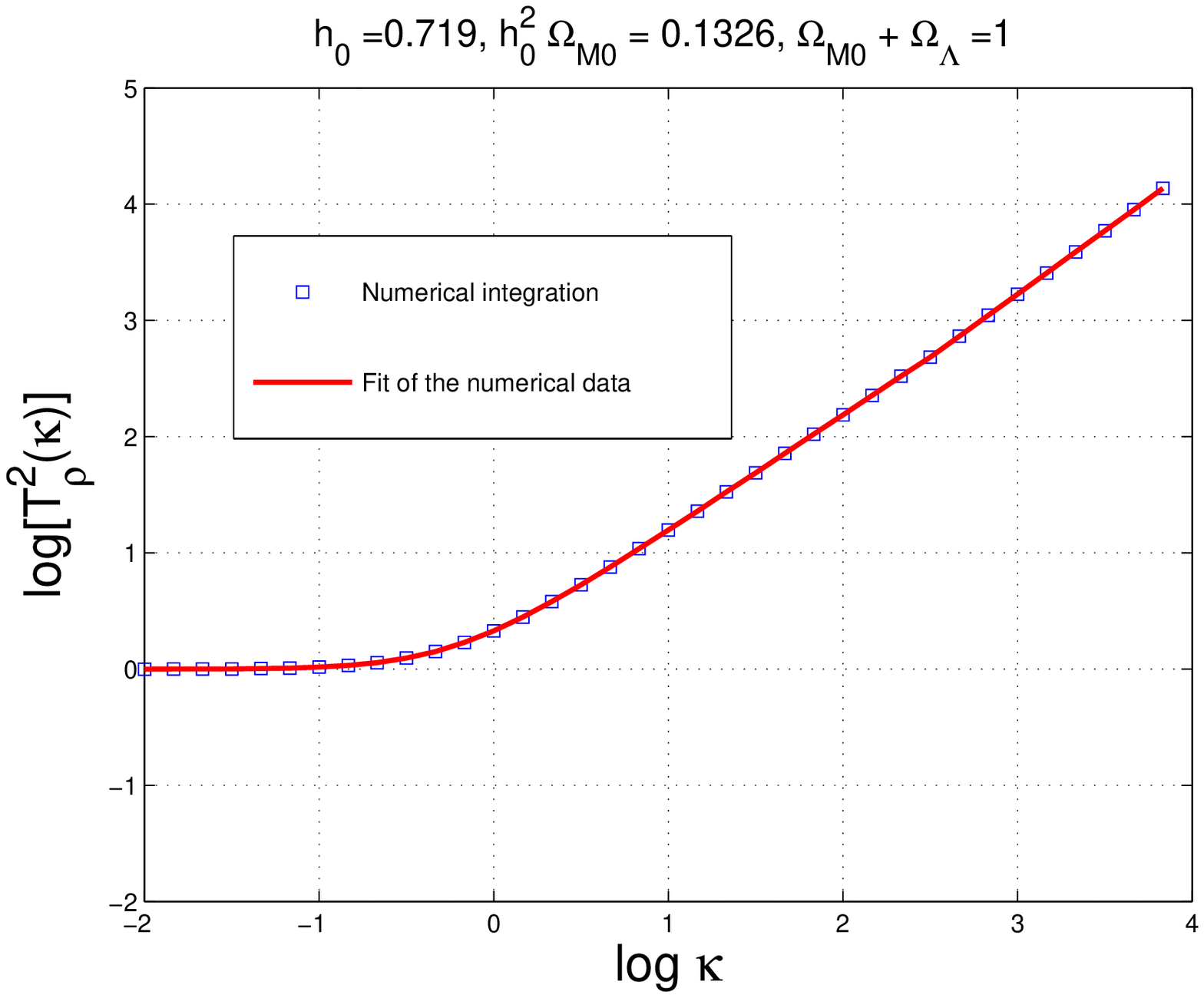}}\\
      \hline
\end{tabular}
\end{center}
\caption{The spectral energy density is integrated across the  transition
between the stiff epoch and the radiation-dominated epoch  
for different values of $\kappa$ (plot at the left). Following the same procedure 
used in the case of Fig. \ref{FIGURE1} the energy transfer function can be obtained and fitted by the appropriate analytic expression (plot at the right).} 
\label{FIGURE2}
\end{figure}
The considerations developed in the case of the radiation-matter transition also apply, for instance, to the stiff-radiation 
transition. 
In Fig. \ref{FIGURE2}, for instance, the transition between a radiation-dominated phase and a stiff phase (with 
$w_{\mathrm{t}}=1$) is illustrated. This time the energy transfer function will be increasing  with the wavenumber (see 
Fig. \ref{FIGURE2}, plot at the right) and the energy transfer function will be given, this time, by
\begin{equation}
T^2(k/k_{\mathrm{s}}) = 1.0  + 0.204\,\biggl(\frac{k}{k_{\mathrm{s}}}\biggr)^{1/4} - 0.980 \,\biggl(\frac{k}{k_{\mathrm{s}}}\biggr)^{1/2}  + 3.389 \biggl(\frac{k}{k_{\mathrm{s}}}\biggr) -0.067\,\biggl(\frac{k}{k_{\mathrm{s}}}\biggr)\ln^2{(k/k_{\mathrm{s}})},
\label{EQ27}
\end{equation}
where $k_{\mathrm{s}} = \tau_{\mathrm{s}}^{-1}$ and $\tau_{\mathrm{s}}$ the time at which the plasma 
becomes dominated by radiation. The fact that the spectral energy density increases linearly (up to logarithmic corrections) fits with the analytical results of \cite{MG2,MG3} where, however, the 
slow-roll corrections were neglected. Further details on this approach will be given in a forthcoming paper \cite{MG5}.

\begin{figure}
\begin{center}
\begin{tabular}{|c|c|}
      \hline
      \hbox{\epsfxsize = 7.2 cm  \epsffile{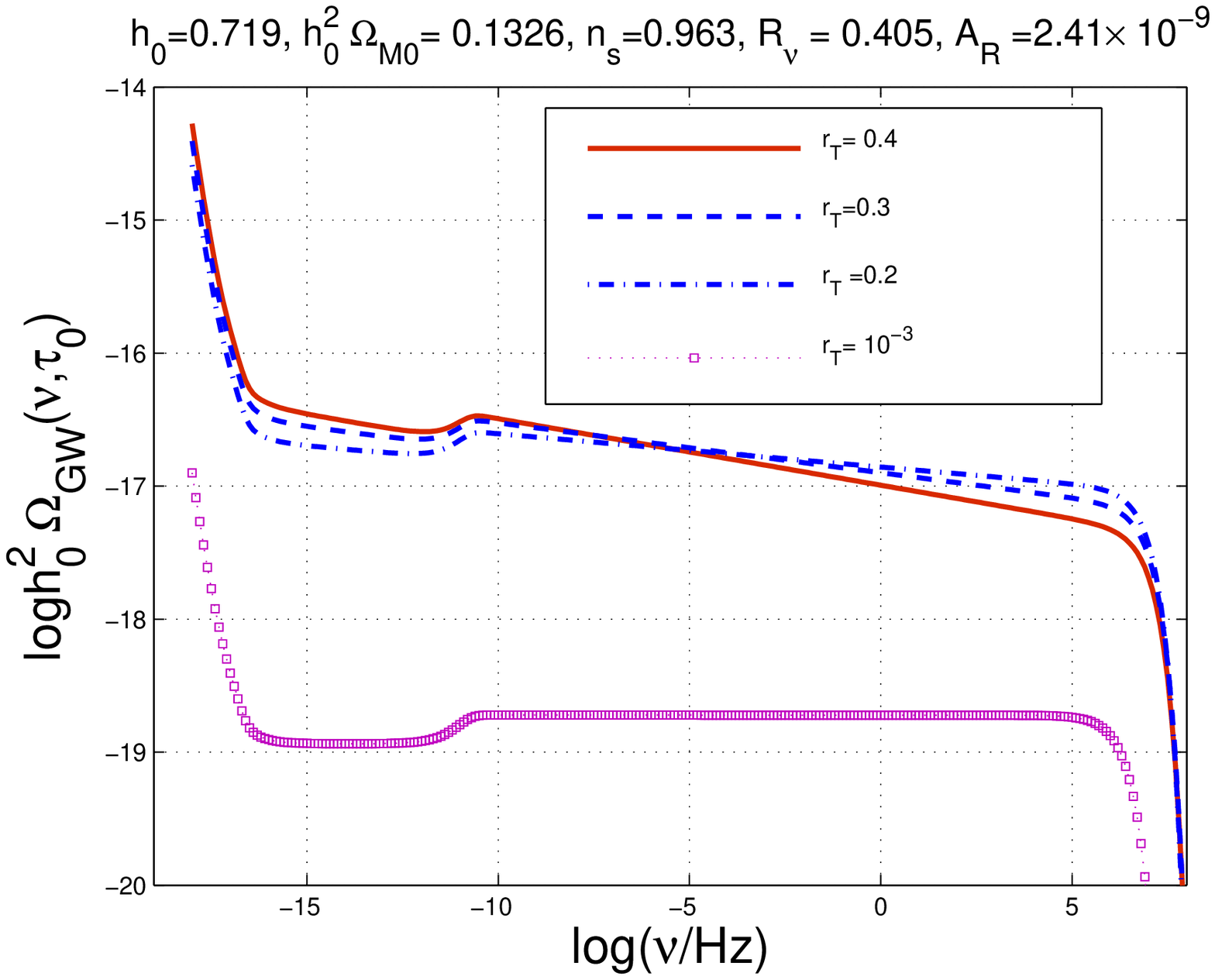}} &
      \hbox{\epsfxsize = 7.2 cm  \epsffile{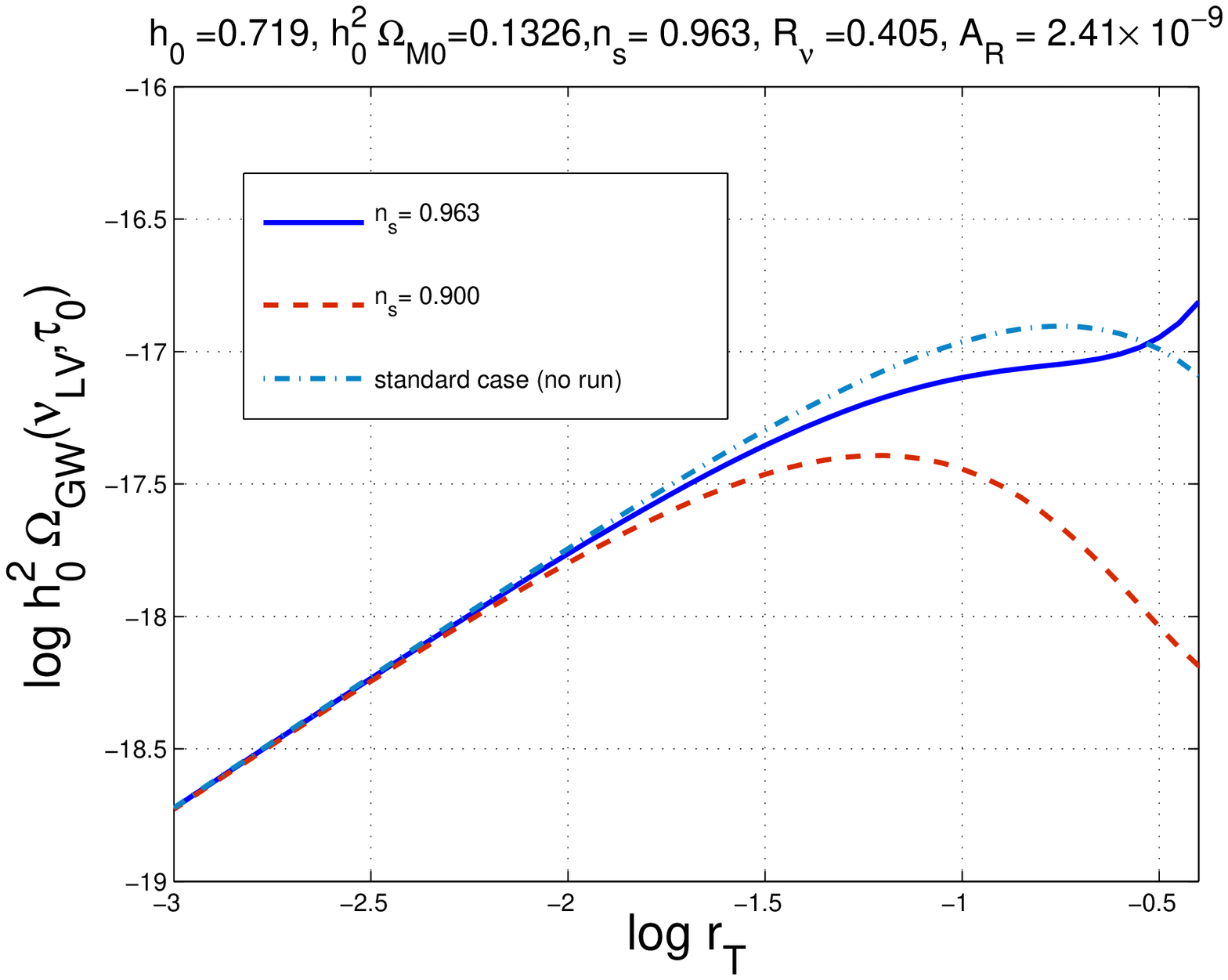}}\\
      \hline
\end{tabular}
\end{center}
\caption{The spectral energy density of the relic gravitons is illustrated in the case of the (conventional) 
$\Lambda$CDM paradigm supplemented by the tensor to scalar ratio $r_{\mathrm{T}}$. The parameters are fixed  
to the best-fit values derived by comparing the $\Lambda$CDM paradigm with the WMAP 5-yr alone \cite{WMAP51,WMAP52}.} 
\label{FIGURE3}
\end{figure}

\begin{figure}
\begin{center}
\begin{tabular}{|c|c|}
      \hline
      \hbox{\epsfxsize = 7.2 cm  \epsffile{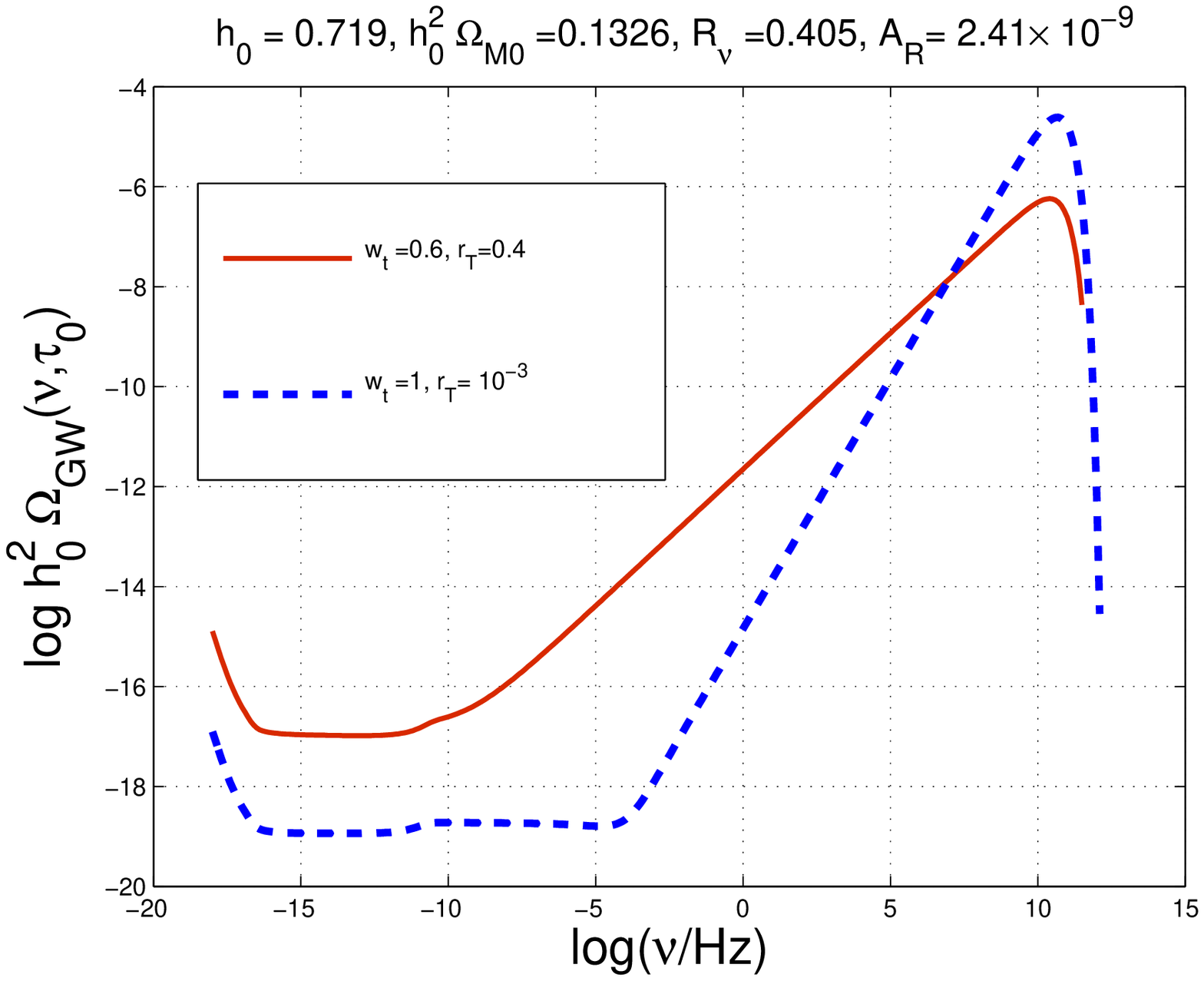}} &
      \hbox{\epsfxsize = 7.2 cm  \epsffile{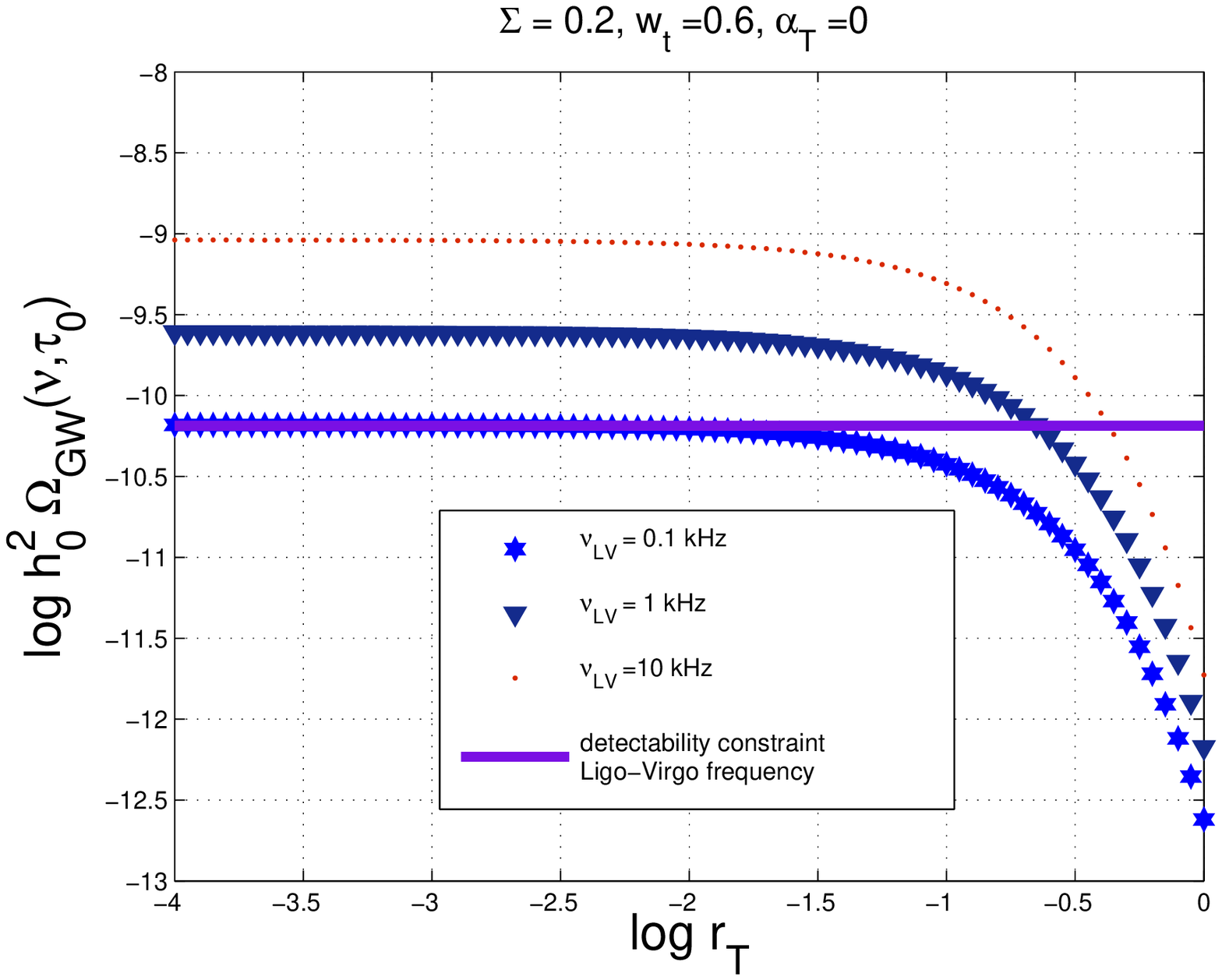}}\\
      \hline
\end{tabular}
\end{center}
\caption{The spectral energy density of the relic gravitons in the case of the 
T$\Lambda$CDM scenario. The parameters are fixed  to the best-fit values derived by comparing the 
$\Lambda$CDM paradigm with the WMAP 5-yr alone \cite{WMAP51,WMAP52}.} 
\label{FIGURE4}
\end{figure}

The outlined  computational procedure allows for a reasonably accurate estimate of the spectral energy density of the relic gravitons in a variety of models. 
In Figs. \ref{FIGURE3} and \ref{FIGURE4}
the spectral energy density is reported, respectively, in the conventional case 
and in the T$\Lambda$CDM scenario taking into account, in both cases, the 
late-time effects which can marginally reduce the amplitude.
Depending upon $R_{\nu}$ (i.e. the neutrino fraction in the radiation plasma), the tensor 
amplitude and the spectral energy density get reduced. For three families of massless neutrinos (as implied by 
the WMAP 5-yr best fits and as assumed in the pivotal $\Lambda$CDM paradigm) 
$R_{\nu} =0.405$ and the amount of suppression is, 
approximately, $0.64$ of the value $\Omega_{\mathrm{GW}}(\nu,\tau_{0})$  has when the very same effect is not taken into account.

The effect of a progressive reduction of relativistic degrees of freedom has been approximately taken into account. In the least favourable case the reduction 
of the relativistic degrees of freedom is flat in frequency and proportional to $(g_{\rho}/g_{\rho0})(g_{\mathrm{s}}/g_{\mathrm{s}0})^{-4/3}$
where $g_{\rho0} =3.36 $,$g_{\mathrm{s}0} =3.90$ \cite{TR1,TR2}. 
Note that $g_{\rho}$ and $g_{\mathrm{s}}$ are the relativistic degrees of freedom 
appearing, respectively, in the energy and in the entropy density.
Finally, there is a modification in the spectrum connected with the late dominance of the 
dark energy \cite{TR1}. The most prominent effect is independent 
on the frequency:  the spectral energy density is suppressed 
by an extra-factor, i.e. $(\Omega_{\mathrm{M}0}/\Omega_{\Lambda})^2$.  In the case of the WMAP 5-yr data 
alone, the $\Lambda$CDM paradigm gives $\Omega_{\Lambda} =0.742$ and $\Omega_{\mathrm{M}0} =0.258$. Intuitively this means that
that ${\mathcal N}_{\rho}$ (appearing in Eq. (\ref{EQ20})) is further suppressed by a factor ${\mathcal O}(0.120)$. 
In Fig. \ref{FIGURE3} (plot at the left) $h_{0}^2 \Omega_{\mathrm{GW}}$ 
is illustrated as a function of the frequency $\nu = k/(2\pi)$ by taking into account all the late-time 
effects mentioned above.  The pivot frequency $\nu_{\mathrm{p}} = 3.092 \,\,\mathrm{aHz}$ corresponds \footnote{Whenever needed, the prefixes of the International System of units will be consistently adopted: 
$1\mathrm{aHz} = 10^{-18}$Hz, $1\mathrm{fHz} = 10^{-15}$Hz and so on.} 
to the pivot wavenumber of Eq. (\ref{EQ13}).
 The spectral energy density (see Fig. \ref{FIGURE1} plot at the left) consists of a decreasing region (at low frequencies) which is 
followed by a nearly scale-invariant plateau  for frequencies $\nu > \nu_{\mathrm{eq}}$ where
\begin{equation}
\nu_{\mathrm{eq}} = \frac{k_{\mathrm{eq}}}{2 \pi} = 1.281 \times 10^{-17} \biggl(\frac{h_{0}^2 \Omega_{\mathrm{M}0}}{0.1326}\biggr) \biggl(\frac{h_{0}^2 \Omega_{\mathrm{R}0}}{4.15 \times 10^{-5}}\biggr)^{-1/2}\,\, \mathrm{Hz},
\label{EQ28}
\end{equation}
is the frequency corresponding to matter-radiation equality \footnote{
In Eq. (\ref{EQ28}) $\Omega_{\mathrm{M}0}$ and $\Omega_{\mathrm{R}0}$ are, respectively, the critical fractions of matter 
and radiation of the putative $\Lambda$CDM model.}.
 The WMAP 5-yr collaboration \cite{WMAP51,WMAP52}
 give an experimental determination of $k_{\mathrm{eq}}$, (i.e. 
$k_{\mathrm{eq}} = 0.00999^{+0.00028}_{-0.00027}\,\, \mathrm{Mpc}^{-1}$) which is fully compatible with the analytical 
estimate of Eq. (\ref{EQ28}). According to Fig. \ref{FIGURE3}, $h_{0}^2 \Omega_{\mathrm{GW}}(\nu,\tau_{0})$ decreases 
exponentially for $\nu > \nu_{\mathrm{max}}$ where 
\begin{equation}
\nu_{\mathrm{max}}  = 0.346 \,\biggl(\frac{\epsilon}{0.01}\biggr)^{1/4} 
\biggl(\frac{{\mathcal A}_{\mathcal R}}{2.41 \times 10^{-9}}\biggr)^{1/4}
\biggl(\frac{h_{0}^2 \Omega_{\mathrm{R}0}}{4.15 \times 10^{-5}}\biggr)^{1/4} \,\mathrm{GHz}.
\label{EQ29}
\end{equation}
While $\nu_{\mathrm{eq}}$ does not depend upon the specific model, 
$\nu_{\mathrm{max}}$ depends, in principle, from the amount of 
redshift between the end of inflation and the present epoch.  
The shallow depression arising in the nearly scale-invariant plateau of Fig. \ref{FIGURE3} (plot at the left) is due to neutrino free streaming and it is present for  
$\nu_{\mathrm{eq}} < \nu< \nu_{\mathrm{bbn}}$ where 
\begin{equation}
\nu_{\mathrm{bbn}} = 
2.252\times 10^{-11} \biggl(\frac{N_{\mathrm{eff}}}{10.75}\biggr)^{1/4} \biggl(\frac{T_{\mathrm{bbn}}}{\,\,\mathrm{MeV}}\biggr) 
\biggl(\frac{h_{0}^2 \Omega_{\mathrm{R}0}}{4.15 \times 10^{-5}}\biggr)^{1/4}\,\,\mathrm{Hz}\simeq 0.01 \,\,\mathrm{nHz}.
\label{EQ30}
\end{equation}
The frequency band of the terrestrial interferometers \cite{virgo,ligo,tama,geo} ranges between few Hz and $10$ kHz with a maximum in the sensitivity to a stochastic background\footnote{The sensitivity to a given signal depends upon various 
factors. For intermediate frequency the signal to noise ratio is also sensitive to the form of the overlap reduction 
function which depends upon the mutual position and relative orientations of the interferometers. The overlap reduction function effectively cuts-off the integral which defines the signal to noise ratio for a typical 
frequency $\nu\simeq 1/(2 d)$ where $d$ is the separation between the two detectors.} 
 for, approximately, $\nu_{\mathrm{LV}} \simeq 0.1$kHz. Since 
$\nu_{\mathrm{eq}} < \nu_{\mathrm{LV}} < \nu_{\mathrm{max}}$, Fig. \ref{FIGURE1} implies (plot at the left) that 
 $h_{0}^2 \Omega_{\mathrm{GW}}(\nu_{\mathrm{LV}},\tau_{0}) \simeq 10^{-17}$.
  To be even more quantitative, in Fig. \ref{FIGURE3} (plot at the right), 
  $h_{0}^2\Omega_{\mathrm{GW}}(\nu_{\mathrm{LV}},\tau_{0})$ 
 is illustrated as a function of $r_{\mathrm{T}}$.  
 In the same plot, the dot-dashed curve  refers to the standard case discussed in Eq. (\ref{EQ14});
  the full and dashed curves refer instead to the situation where the spectral index depends upon the 
 frequency as $n_{\mathrm{T}} = -r_{\mathrm{T}}/8 + (r_{\mathrm{T}}/16)[(n_{\mathrm{s}} -1) + (r_{\mathrm{T}}/8)] \ln{(\nu/\nu_{\mathrm{p}})}$. Figure \ref{FIGURE3} shows that, in both situations, 
  $h_{0}^2 \Omega_{\mathrm{GW}}(\nu_{\mathrm{LV}},\tau_{0}) \simeq {\mathcal O}10^{-17}$  given the current  limits on $r_{\mathrm{T}}$.
 
In the case of an exactly scale invariant spectrum
the correlation of the two (coaligned) LIGO detectors with 
central corner stations in Livingston (Lousiana) and in Hanford 
(Washington) might reach a sensitivity to a flat spectrum 
which is \cite{MG1}
\begin{equation}
h_0^2\,\, \Omega_{\rm GW}(\nu_{\mathrm{LV}}) \simeq 6.5 \times 10^{-11} \,\, 
\biggl(\frac{1\,\,\mathrm{yr}}{T} \biggr)^{1/2}\,\,\mathrm{SNR}^2, \qquad \nu_{\mathrm{LV}} =0.1 \,\, \mathrm{kHz}
\label{EQ31}
\end{equation} 
where $T$ denotes the observation time and $\mathrm{SNR}$ is the signal to noise ratio.  Equation (\ref{EQ31}) is in close agreement with the 
sensitivity of the advanced Ligo apparatus \cite{ligo} to an exactly scale-invariant spectral energy density \cite{AR}.  Equation (\ref{EQ31}) together with the 
plots at the right in Fig. \ref{FIGURE3} suggest that the relic graviton 
background predicted by the $\Lambda$CDM paradigm is not directly 
observable by wide-band interferometers in their advanced incarnation.
The minuteness of $h_{0}^2 \Omega_{\mathrm{GW}}(\nu_{\mathrm{LV}},\tau_{0})$ stems directly from the 
assumption that the inflationary phase is  suddenly followed by the radiation-dominated phase.

Let us then posit that between the end of  
inflation and the onset of the radiation-dominated phase a sufficiently   long 
stiff phase takes place.  In this case the spectral energy density of the relic gravitons 
will increase for frequencies larger than $\nu_{\mathrm{s}} = k_{\mathrm{s}}/(2 \pi)$. 
Assuming that the inflationary phase is be of quasi-de Sitter type 
and characterized by a given value of $r_{\mathrm{T}}$,  it must always happen, no 
matter how the parameters of the model are assigned, that $\nu_{\mathrm{s}} > \nu_{\mathrm{bbn}}$.

The frequency scale $\nu_{\mathrm{s}}$ is related to the 
duration of the stiff phase and it is bounded from below by the nucleosynthesis frequency. The slope of the spectral energy density in the high-frequency branch is related, ultimately, to the sound speed and it is bounded, from above, by the speed of light. These are the two supplementary  parameters of  T$\Lambda$CDM scenario.
In Fig. \ref{FIGURE4} (see plot at the left)  the spectral energy density 
computed in the T$\Lambda$CDM scenario is illustrated
for two different values of $w_{\mathrm{t}}$ and $r_{\mathrm{T}}$.   For $\nu > \nu_{\mathrm{s}}$ the spectral 
 energy density acquires a blue spectrum \footnote{The spectrum is blue, in general terms, if it is 
 increasing with frequency. The slow-roll dynamics always implies, within the $\Lambda$CDM scenario, 
  much milder scaling violations which involve only red spectra, i.e. spectra which are very slowly decreasing in frequency 
  like those of Fig. \ref{FIGURE1} (see plot at the left).}.
 
Defining as 
$H \simeq (\epsilon \pi {\mathcal A}_{{\mathcal R}})^{1/2} \, M_{\mathrm{P}}$ the typical inflationary 
curvature scale and as  $H_{\mathrm{r}}$ the Hubble rate at the onset of the radiation epoch, $\nu_{\mathrm{s}}$ and 
$\nu_{\mathrm{max}}$ can be written, in the T$\Lambda$CDM scenario, as
\begin{eqnarray}
&&\nu_{\mathrm{s}} = 1.177\times 10^{11} \Sigma^{\gamma} \,\,(\pi \epsilon {\mathcal P}_{{\mathcal R}})^{\frac{\gamma +1}{4}}\,\,\biggl(\frac{h_{0}^2 \Omega_{\mathrm{R}0}}{4.15 \times 10^{-5}}\biggr)^{1/4}\,\,\mathrm{Hz},
\label{EQ32}\\
&& \nu_{\mathrm{max}} = 1.177 \times 10^{11} \Sigma^{-1} 
\biggl(\frac{h_{0}^2 \Omega_{\mathrm{R}0}}{4.15 \times 10^{-5}}\biggr)^{1/4}\,\,\mathrm{Hz}, \qquad 
\Sigma =\biggl(\frac{{\it H}}{M_{\mathrm{P}}}\biggr)^{\frac{\gamma +1}{2 \gamma}}
 \biggr(\frac{{\it H}_{\mathrm{r}}}{M_{\mathrm{P}}}\biggl)^{\frac{1}{2 \gamma}},
\label{EQ33}
\end{eqnarray}
where $\gamma\equiv \gamma(w_{\mathrm{t}}) = 3(w_{\mathrm{t}} +1)/(3 w_{\mathrm{t}} -1)$. By definition $\Sigma$ 
is fully determined by fixing $H_{\mathrm{r}}$. So, 
 $\Sigma$ and $\gamma(w_{\mathrm{t}})$ can be chosen as the two 
pivotal parameters of the T$\Lambda$CDM scenario. Equivalently 
$\Sigma$ and $\gamma$ can be traded for $\nu_{\mathrm{s}}$ and 
for the slope of the spectral energy density during the stiff phase (which is given, up to logarithmic corrections) by 
$(6 w_{\mathrm{t}} -2)/[(3 w_{\mathrm{t}} +1)]$. As stressed above, the natural upper limit for the spectral slope is exactly $1$ which is 
the maximally stiff fluid compatible with causality \cite{EMM}. 

The frequency $\nu_{\mathrm{s}}$ can be much larger than $\nu_{\mathrm{bbn}}$ (for instance 
$\nu \simeq \mathrm{mHz}$ in \cite{MG3}) but cannot be smaller than $\nu_{\mathrm{bbn}}$ which constitutes a natural lower limit for $\nu_{\mathrm{s}}$.  If $\nu_{\mathrm{s}} < \nu_{\mathrm{bbn}}$ the plasma would be stiff 
also throughout nucleosynthesis which is unacceptable.
The observed abundances of the light elements (together with CMB data) also constrain the total energy density of the relic gravitons, i.e. the integral of 
$\Omega_{\mathrm{GW}}(\nu,\tau_{0})$ over the frequency.  This bound is usually expressed as\footnote{Coherently with established conventions $\ln$ will denote 
the natural logarithm, while the logarithms to base $10$ (i.e. common logarithms) will 
be denoted by $\log$.}:
\begin{equation}
h_{0}^2 \Omega_{\mathrm{GW}}(\tau_{0}) =h_{0}^2  \int_{\nu_{\mathrm{bbn}}}^{\nu_{\mathrm{max}}}
  \Omega_{{\rm GW}}(\nu,\tau_{0}) d\ln{\nu} = 5.6 \times 10^{-6} 
  \biggl(\frac{h_{0}^2 \Omega_{\gamma0}}{2.47 \times 10^{-5}}\biggr)\Delta N_{\nu} 
\label{EQ34}
\end{equation}
where $\Delta N_{\nu}$ is the equivalent number of extra-relativistic species at the onset of standard big-bang 
nucleosynthesis\footnote{The language of Eq. (\ref{EQ34}) may seem a bit contrived but it is a simple 
consequence of the historical development of the field. The extra-relativistic species 
were associated, in the past, with families of neutrinos. The nature of the bound on $\Delta N_{\nu}$ (and hence 
on $h_{0}^2 \Omega_{\mathrm{GW}}(\tau_{0})$) does not 
change if the relativistic species are bosonic (like in the case of gravitons). For a discussion of the derivation
of Eq. (\ref{EQ8}) see \cite{MG4}.}. 
In the standard scenario for the synthesis of light nuclei, 
$0.2 <\Delta N_{\nu} <1$ and, therefore $h_{0}^2 \Omega_{\mathrm{GW}}(\tau_{0}) $ will be constrained accordingly.  In Fig. \ref{FIGURE2} (plot at the right) the spectral energy density is reported as function of $r_{\mathrm{T}}$ in the context of the T$\Lambda$CDM scenario 
and for typical frequencies in the operating window of wide-band interferometers.  
As $r_{\mathrm{T}}$ diminishes,
the amplitude of the spectral energy density is almost constant. 
The latter occurrence arises for 
two independent reasons. On one hand the most relevant constraint, in the case of growing spectral energy densities, is the one 
provided by Eq. (\ref{EQ8}) and enforced in both plots of Fig. \ref{FIGURE2}. On the other hand 
the frequency $\nu_{\mathrm{s}}$ depends also upon $r_{\mathrm{T}}$ (through $\epsilon$, see Eqs (\ref{EQ14}) and (\ref{EQ33})).
It should be finally appreciated,  from Figs. \ref{FIGURE2} and \ref{FIGURE3}, that the pulsar timing bounds 
(recently revisited \cite{PUL}) still imply that $h_{0}^2 \Omega_{\mathrm{GW}}(\nu_{\mathrm{pulsar}},\tau_{0}) < 1.9 \times 
10^{-8}$ for a $\nu_{\mathrm{pulsar}} \simeq 10$ nHz which is roughly comparable with the inverse 
of the observation time along which the pulsars timing has been monitored. Such a bound is not constraining  for the T$\Lambda$CDM model. The proof goes as follows. Assuming the maximal growth of the spectral energy density (i.e. that $h_{0}^2 \Omega_{\mathrm{GW}}(\nu,\tau_{0}) \propto \nu$)  and the minimal value of $\nu_{\mathrm{s}}$
(i.e. $\nu > \nu_{\mathrm{bbn}}$), we will have 
that, at the frequency scale of the pulsars,  $h_{0}^2\Omega_{\mathrm{GW}}(\nu_{\mathrm{pulsar}},\tau_{0}) \simeq 10^{-13}$  or even $10^{-14}$ depending upon 
$r_{\mathrm{T}}$.  But this value is always much smaller than the constraint 
stemming from pulsar timing measurements.

In this paper it has been suggested that the $\Lambda$CDM parameter can be 
complemented by adding a post-inflationary phase characterized by a sound speed larger than the one of an ultra-relativistic plasma (i.e. $1/\sqrt{3}$). Causality constrains the maximal 
barotropic index and the maximal sound speed. Big bang nucleosynthesis sets limits 
both on  the maximal duration of the stiff phase and on the total energy density of the 
relic gravitons.
Two new parameters will then be added to the $\Lambda$CDM paradigm which has been dubbed, throughout the paper, as 
tensor- $\Lambda$CDM (T$\Lambda$CDM) paradigm since relic gravitons are copiously produced at high frequencies (i.e. larger than $0.1$ nHz).  
The new pivot frequency defines the scale at which the spectral energy density of the relic gravitons 
starts increasing with a slope which is dictated by the stiff barotropic index.  In the T$\Lambda$CDM scenario, 
which may be seen as an improved version of the models proposed in \cite{MG2,MG3}, the spectral energy density of the relic gravitons can even  be from $6$ to $7$ orders of magnitude larger than in conventional inflationary models.
Along a more technical perspective, a numerical recipe for the calculation of the spectral energy density has been presented.  

The advanced versions of wide-band interferometers are germane to the theme 
of the present investigation.  
 At the moment the CMB data \cite{WMAP51,WMAP52}, 
large-scale structure observations \cite{LSS1,LSS2} and supernovae light curves \cite{SN1,SN2}
are used in combined analysis to put bounds on $r_{\mathrm{T}}$, i.e. the tensor to scalar ratio. Few years from now the three  aforementioned cosmological data sets will still be used to constrain (and hopefully determine) 
$r_{\mathrm{T}}$ while, given the foreseen sensitivities, the (terrestrial) wide-band interferometers will still 
be unable to set concurrent limits to backgrounds of relic gravitons. Provided the claimed sensitivities will be  
reached in due time, the considerations presented here give a concrete opportunity of using interferometers data together 
with the more classic cosmological data sets to rule out (or, more optimistically, rule in) a class of specific models.  
It is productive to stress that, in the present context, any potential upper limit from wide-band interferometers 
will directly constrain the post-inflationary thermal history. 
Cosmology is not tested in a laboratory: therefore the nature of the observations is
 inextricably bound to the models employed to analyze the data and to the potential 
 redundancy of different data sets.  It has been shown here that by 
 complementing a known model with supplementary physical considerations,  
 the three established cosmological data sets can also profit of a qualitatively new 
 class of observations, such as the ones provided by wide-band interferometers. 
It is tempting to speculate that the perspective of the present paper could provoke a useful synergy between communities scrutinizing different branches of the graviton spectrum.  
The fruitful dialogue between the experiments sensitive to small (i.e. $\nu_{\mathrm{p}} \simeq \,\mathrm{aHz}$) and to intermediate frequencies (i.e. $\nu_{\mathrm{LV}}\simeq 0.1$kHz) could be extended, in principle, also to conceptually different kinds of 
detectors such as microwave cavities \cite{CV1} and  waveguides \cite{CV2}.
 
It is a pleasure to acknowledge interesting discussions with E. Picasso.


\begin{thebibliography}{99}

\bibitem{WMAP51}  G.~Hinshaw {\it et al.}  [WMAP Collaboration],  arXiv:0803.0732 [astro-ph];  J.~Dunkley {\it et al.}  [WMAP Collaboration], arXiv:0803.0586 [astro-ph].

\bibitem{WMAP52} B.~Gold {\it et al.}  [WMAP Collaboration], arXiv:0803.0715 [astro-ph];  E.~Komatsu {\it et al.}  [WMAP Collaboration],  arXiv:0803.0547 [astro-ph];
M.~Nolta {\it et al.}  [WMAP Collaboration],  arXiv:0803.0593 [astro-ph].

\bibitem{LSS1} W.~L.~Freedman {\it et al.}, Astrophys.\ J.\  {\bf 553}, 47 (2001);
S.~Cole {\it et al.}  [The 2dFGRS Collaboration],  Mon.\ Not.\ Roy.\ Astron.\ Soc.\  {\bf 362} , 505 (2005);  

\bibitem{LSS2} D.~J.~Eisenstein {\it et al.}  [SDSS Collaboration], Astrophys.\ J.\  {\bf 633}, 560 (2005); M.~Tegmark {\it et al.}  [SDSS Collaboration], Astrophys.\ J.\  {\bf 606}, 702 (2004).

\bibitem{SN1} P.~Astier {\it et al.}  [The SNLS Collaboration], Astron.\ Astrophys.\  {\bf 447}, 31 (2006).

\bibitem{SN2} A.~G.~Riess {\it et al.}  [Supernova Search Team Collaboration],  Astrophys.\ J.\  {\bf 607}, 665 (2004); B.~J.~Barris {\it et al.}, Astrophys.\ J.\  {\bf 602}, 571 (2004).

 \bibitem{ligo}  A. Abramovici et al., Science {\bf 256}, 325 (1992); http://www.ligo.org
 
\bibitem{virgo}  B. Caron et al., Class. Quant. Grav. {\bf 14}, 1461 (1997); http://www.virgo.infn.it
 
 \bibitem{tama}  M. Ando et al., Phys. Rev. Lett. {\bf 86}, 3950 (2001); http://tamago.mtk.nao.ac.jp
  
 \bibitem{geo}  H. L\"uck et al., Class. Quant. Grav. {\bf 14}, 1471 (1997); http://www.geo600.uni-hannover.de

\bibitem{stoch1}  B.~Abbott {\it et al.}  [ALLEGRO Collaboration and LIGO Scientific
  Collaboration], Phys.\ Rev.\  D {\bf 76}, 022001 (2007);   G.~Cella, {\it et al.}  Class.\ Quant.\ Grav.\  {\bf 24}, S639 (2007);
  L.~Baggio {\it et al.}  [AURIGA Collaboration],
  Class.\ Quant.\ Grav.\  {\bf 25}, 095004 (2008).
  
\bibitem{THTool} M. Giovannini,  Int.\ J.\ Mod.\ Phys.\ D  {\bf 13}, 391 (2004).

\bibitem{efolds}  A.~R.~Liddle and S.~M.~Leach,  Phys.\ Rev.\  D {\bf 68}, 103503 (2003).
 
 \bibitem{MG2} M.~Giovannini,  Phys.\ Rev.\  D {\bf 58}, 083504 (1998).

\bibitem{MG3} M.~Giovannini, Class.\ Quant.\ Grav.\  {\bf 16}, 2905 (1999);  Phys.\ Rev.\  D {\bf 60}, 123511 (1999).

\bibitem{ZEL1}  Y.~B.~Zeldovich,  Mon.\ Not.\ Roy.\ Astron.\ Soc.\  {\bf 160}, 1P (1972).

\bibitem{PV1}  P.~J.~E.~Peebles and A.~Vilenkin,  Phys.\ Rev.\  D {\bf 59}, 063505 (1999).

\bibitem{REP}   V.~Sahni, M.~Sami and T.~Souradeep, Phys.\ Rev.\  D {\bf 65}, 023518 (2002);  H.~Tashiro, T.~Chiba and M.~Sasaki, Class.\ Quant.\ Grav.\  {\bf 21}, 1761 (2004); 
 T.~J.~Battefeld and D.~A.~Easson,  Phys.\ Rev.\  D {\bf 70}, 103516 (2004).

\bibitem{EMM}   G.~Ellis, R.~Maartens and M.~A.~H.~MacCallum, Gen.\ Rel.\ Grav.\  {\bf 39}, 1651 (2007).
  
 \bibitem{nu1} S.~Weinberg, Phys.\ Rev.\  D {\bf 69}, 023503 (2004); D.~A.~Dicus and W.~W.~Repko,  Phys.\ Rev.\  D {\bf 72}, 088302 (2005).
  
\bibitem{TR1}   W.~Zhao and Y.~Zhang, Phys.\ Rev.\  D {\bf 74}, 043503 (2006); Y.~Zhang, W.~Zhao, T.~Xia and Y.~Yuan, Phys.\ Rev.\  D {\bf 74}, 083006 (2006).

\bibitem{TR2} Y.~Watanabe and E.~Komatsu,  Phys.\ Rev.\  D {\bf 73}, 123515 (2006).

\bibitem{TR3}  S.~Chongchitnan and G.~Efstathiou, Phys.\ Rev.\  D {\bf 73}, 083511 (2006); Prog.\ Theor.\ Phys.\ Suppl.\  {\bf 163}, 204 (2006).

 \bibitem{ford} L. H. Ford and L. Parker, Phys. Rev. D {\bf 16},1601 (1977);   Phys. Rev. D {\bf 16}, 245  (1977).

\bibitem{isaacson} R. Isaacson, Phys. Rev. {\bf 166}, 1263 (1968); Phys. Rev. {\bf 166}, 1272 (1968).

\bibitem{BR}  M.~Giovannini,  Phys.\ Rev.\  D {\bf 73}, 083505 (2006);  L.~Abramo, Phys.\ Rev.\  D {\bf 60}, 064004 (1999).

\bibitem{FS1}   M.~S.~Turner, M.~J.~White and J.~E.~Lidsey,  Phys.\ Rev.\  D {\bf 48}, 4613 (1993).

\bibitem{MG5} M.~Giovannini, in preparation.

\bibitem{MG1}   D.~Babusci and M.~Giovannini, Phys.\ Rev.\  D {\bf 60}, 083511 (1999); Class.\ Quant.\ Grav.\  {\bf 17}, 2621 (2000); 
  Int.\ J.\ Mod.\ Phys.\  D {\bf 10}, 477 (2001).

\bibitem{AR} B.~Allen and J.~D.~Romano,  Phys.\ Rev.\  D {\bf 59}, 102001 (1999).

\bibitem{MG4} M.~Giovannini, H.~Kurki-Suonio and E.~Sihvola, Phys.\ Rev.\  D {\bf 66}, 043504 (2002).

\bibitem{PUL} F.~A.~Jenet {\it et al.}, Astrophys.\ J.\ {\bf 653}, 1571 (2006) [arXiv:astro-ph/0609013].
  
 \bibitem{CV1}  F. Pegoraro, L. A. Radicati, Ph. Bernard, and E. Picasso,
Phys. Lett. A {\bf 68}, 165 (1978); Ph.~Bernard, G.~Gemme, R.~Parodi and E.~Picasso,  Rev.\ Sci.\ Instrum.\  {\bf 72}, 2428 (2001);
 R.~Ballantini, P.~Bernard, A.~Chincarini, G.~Gemme, R.~Parodi and E.~Picasso,
  Class.\ Quant.\ Grav.\  {\bf 21}, S1241 (2004).
 
\bibitem{CV2}  A. M. Cruise, Class. Quantum Grav. {\bf 17} , 2525 (2000);  A. M. Cruise and R. M. Ingley, Class. \ Quantum \ Grav. {\bf 23},  6185 (2006);
F.~Y.~Li, M.~X.~Tang and D.~P.~Shi,  Phys.\ Rev.\  D {\bf 67}, 104008 (2003).

\end{thebibliography}
\end{document}